\documentclass[aps,pra,reprint,floatfi]{revtex4-1}

\usepackage[utf8]{inputenc}
\usepackage{graphicx}
\usepackage{amsmath}
\usepackage{amsfonts}
\usepackage{amssymb}
\usepackage{bbold}
\usepackage{mathptmx}
\DeclareMathAlphabet{\mathcal}{OMS}{cmsy}{m}{n}

\begin{document}

\title{Typical growth behavior of the out-of-time-ordered commutator in many-body localized systems}

\author{Juhee Lee}
\affiliation{Department of Physics and Photon Science, School of Physics and Chemistry, Gwangju Institute of Science and Technology, Gwangju 61005, Korea}
\author{Dongkyu Kim}
\affiliation{Department of Physics and Photon Science, School of Physics and Chemistry, Gwangju Institute of Science and Technology, Gwangju 61005, Korea}
\author{Dong-Hee Kim}
\email{dongheekim@gist.ac.kr}
\affiliation{Department of Physics and Photon Science, School of Physics and Chemistry, Gwangju Institute of Science and Technology, Gwangju 61005, Korea}

\begin{abstract}
We investigate the typicality of the growth behavior of the out-of-time-ordered commutator (OTOC) in the many-body localized (MBL) quantum spin chains across random disorder realizations. In the MBL phase of the Heisenberg XXZ chain, we find that the estimate of the OTOC fluctuates significantly with the disorder realizations at the intermediate times of the main growth. Despite the consequent failure of the disorder average in the MBL phase, we argue that the characteristic behavior of the OTOC growth can still be identified by going through individual disorder realizations. We find that a power-law-type growth behavior appears typically after a disorder-dependent relaxation period, which is very close to the $t^2$ form derived in the effective Hamiltonian of a fully MBL system. The characteristic growth behavior observed at an individual disorder realization is robust in our tests with various state preparations and also verified in another MBL system of the random-transverse-field quantum Ising chain in a uniform longitudinal field.
\end{abstract}

\maketitle

\section{Introduction}

The out-of-time-ordered (OTO) commutator and correlator \cite{Larkin1968,Kitaev2014} 
have attracted much attention recently because of their promising applications
to the diagnosis of information scrambling
\cite{Page1993,Hayden2007,Sekino2008,Lashkari2013}
in the dynamics of quantum many-body systems.
For two unitary operators $\hat{W}$ and $\hat{V}$ that are initially local 
at different positions, the OTO commutator is defined as the expectation value of 
their squared commutator,
\begin{equation} \label{eq:C}
C(t) = \frac{1}{2} \big\langle  [ \hat{W}(t),\hat{V} ]^\dagger 
[ \hat{W}(t),\hat{V} ] \big\rangle \, ,
\end{equation}
which is often rewritten as $C(t) = 1 - \mathrm{Re}[F(t)]$ in terms of
the corresponding OTO commutator
$F(t) = \langle \hat{W}^\dagger(t) \hat{V}^\dagger \hat{W}(t) \hat{V} \rangle$. 
The evaluation of $\langle \cdots \rangle \equiv \mathrm{Tr}[\hat{\rho} \cdots]$ 
depends on the density matrix $\hat{\rho}$ of a pure or mixed state 
prepared for measurement. 
While the two distant operators commute with each other initially,
$\hat{W}(t)$ in time evolution can be highly non-local 
in the presence of interactions, breaking the initial commutativity 
due to an overlap with $\hat{V}$.
The growth of $C(t)$ can thus quantify such scrambling of information 
that spreads across nonlocal degrees of freedom.
The characterization of quantum dynamics by using the OTO commutator 
and correlator has been the subject of intense study in systems ranging from 
a black hole \cite{Shenker2014,Roberts2015,Cotler2017,Grozdanov2018} 
and Sachdev-Ye-Kitaev-type models \cite{Fu2016,Maldacena2016a,Gu2017,Banerjee2017,Lucas2017}
to various condensed matter models of chaotic
\cite{Maldacena2016b,Scaffidi,Patel2017a,Patel2017b,Stanford2016,Chowdhury2017,Shen2017,Rozenbaum2017,Bohrdt2017,Kukuljan2017,Luitz2017,Syzranov2018,Xu,Zhang2019,Khemani2018,Rammensee2018,Garcia-Mata2018,Alvirad2019}
and non-chaotic 
\cite{Dora2017,Tsuji2017,Lin2018,TorresHerrera2018,TorresHerrera2018,Syzranov2019,McGinley2019,Fan2017,He2017,Swingle2017,Chen2017,Chen,He2017,Chen2017,Huang2017,Slagle2017,Bordia2018,Riddell2019} 
systems.
Measurement protocols have been proposed 
\cite{Swingle2016,Yao,Zhu2016,YungerHalpern2017,YungerHalpern2018,Dressel2018}, 
and there are recent advances in experimental measurements 
using nuclear spins of molecules~\cite{Li2017}, trapped ions~\cite{Garttner2017}, 
and ultracold gases~\cite{Meier}.

One of the fundamental questions on the growth behavior of the OTO commutator may be 
whether there exists a characteristic form that can distinguish systems between 
different classes of information scrambling. In a chaotic system, $C(t)$ 
grows very fast, which is often described by the exponential behavior 
with the Lyapunov exponent. In the absence of chaos, the growth of $C(t)$ 
can be much slower or even absent. In particular, measuring $C(t)$ in disordered 
systems may distinguish 
many-body localization~\cite{Basko2006,Oganesyan2007,Nandkishore2015,Altman2015} 
from Anderson localization~\cite{Anderson1958}.
Both of them arrest particle transport, but in many-body localized (MBL)
systems, the dephasing effects due to the interactions allow the spreading of 
quantum information, leading to characteristic dynamics 
\cite{Serbyn2014a,Serbyn2014b,Vasseur2015,Serbyn2017}
including slowly growing $C(t)$
\cite{Fan2017,He2017,Swingle2017,Chen2017,Chen,He2017,Chen2017,Huang2017,Slagle2017,Bordia2018}. 

The specific question that we want to address here is how typical a particular 
slow-growth behavior of $C(t)$ is across different disorder realizations
in a disordered MBL system. In the effective ``l-bit'' Hamiltonian of 
a fully MBL system, it was shown that the growth of the disorder average
$\overline{C(t)}$ proceeds as $\overline{C(t)} \sim t^2$ 
at early times, and then it is saturated with a power-law decaying 
second moment at late times~\cite{Fan2017,Swingle2017,Chen2017}. 
However, beyond the effective Hamiltonian, it is not entirely clear 
how universal this particular power-law form of the growth is 
in more realistic models and whether or not it can characterize 
a wider range of the disordered MBL systems.

In this paper, we investigate the time evolution of $C(t)$ in two 
quantum spin models of the Heisenberg XXZ and mixed-field Ising chains 
across their random disorder realizations. We find that in the MBL phase, 
the disorder average cannot properly show the behavior of $C(t)$ 
because of the large deviations across the disorder realizations. 
The estimate of $C(t)$ at an intermediate time exhibits a bimodal distribution, 
which is in contrast to a singly peaked distribution evolving 
in the ergodic side. 
However, by going through individual disorder realizations, we find that 
the systems typically undergo common stages of the OTO commutator growth. 
At very early times, $C(t)$ shows an intrinsic power-law growth, 
and it is shortly relaxed by the period of an oscillatory plateau that makes 
an offset at a small value of $C$ for the main stage of the growth emerging 
at intermediate times.

For an individual disorder realization, the main growth of $C(t)$ turns out 
to be often very close to the $t^2$ form, while it appears with an offset 
and a time delay that fluctuate very much with disorder realizations.
Based on numerical observations, we argue that the growth can be 
characterized as $C(t) \sim c_0 + \epsilon t^2$ with an offset $c_0$ 
at intermediate times by combining the coarse-grained effect of 
fast dynamics and the slow $t^2$ contribution emerging from 
the effective Hamiltonian.
We also observe that the growth behavior at a fixed disorder realization 
is insensitive to the choice of the state $\hat{\rho}$ that we examine 
for the measurement of $C(t)$. 

This paper is organized as follows. In Sec.~\ref{sec:lbit}, we revisit 
the effective l-bit Hamiltonian for the OTO commutator growth 
behavior at a given disorder configuration. In Sec.~\ref{sec:XXZ}, 
we present the typicality and deviations of the growth behavior 
across random disorder realizations in the Heisenberg XXZ model. 
The representativeness of disorder average is examined,
and the growth behavior is characterized at an individual disorder realization.
In Sec.~\ref{sec:Ising}, we verify the appearance of the characteristic 
growth form in the mixed-field Ising model. 
The summary and conclusions are given in Sec.~\ref{sec:summary}.

\section{The phenomenological ``l-bit'' model}
\label{sec:lbit}

Let us briefly revisit the behavior of the OTO commutator 
in the phenomenological model of a fully MBL system. 
The effective ``l-bit'' Hamiltonian \cite{Serbvn2013,Huse2014,Imbrie2016} 
is written as
\begin{equation} \label{eq:Heff}
\mathcal{H} = \sum_i h_i \hat{\tau}^z_i 
+ \sum_{\{i,j\}} J_{ij} \hat{\tau}^z_i \hat{\tau}^z_j  
+ \sum_{\{i,j,k\}} K_{ijk} \hat{\tau}^z_i \hat{\tau}^z_j \hat{\tau}^z_k 
+ \cdots \, ,
\end{equation}
where $\hat{\tau}^z_i$ is the $z$ component of the Pauli operator 
for a dressed spin-$1/2$ localized at site $i \in [1,L]$ 
in the chain of length $L$. The summation runs over 
the sets of unique site indices. 
The coefficients are given as random variables for the multispin 
interactions which have characteristic strength decaying exponentially 
with distance between the farthest-apart spins. 
The OTO correlator $F(t)$ was derived previously for the operator choice of 
$\hat{W} = \hat{\tau}^x_a$ and 
$\hat{V} = \hat{\tau}^x_b$ \cite{Fan2017,Swingle2017,Chen2017}. 
From the previous results, we can write down the corresponding 
OTO commutator $C(t)$ for a given disorder realization as
\begin{eqnarray} \label{eq:C_lbit}
C(t) &=& 1 - \mathrm{Re} \big[ \big\langle 
\exp \big(it \, 4\hat{J}^{\mathrm{eff}}_{ab} 
\hat{\tau}^z_a \hat{\tau}^z_b \big) \big\rangle \big]
\\ \nonumber
& \simeq & 1 - \cos \big( 4t\langle \hat{J}^{\mathrm{eff}}_{ab}\rangle \big)
\exp \left[ -8t^2 \big( \langle [\hat{J}^{\mathrm{eff}}_{ab}]^2 \rangle 
- \langle \hat{J}^{\mathrm{eff}}_{ab}\rangle^2 \big) \right] \, .
\end{eqnarray}
The effective interaction operator $\hat{J}^{\mathrm{eff}}_{ab}$ for the spins 
at $a$ and $b$ can be written by collecting all terms involving $a$ and $b$ as
\begin{equation} \label{eq:Jeff}
\hat{J}^{\mathrm{eff}}_{ab} = J_{ab} 
+ \sideset{}{'}\sum_k K_{abk} \hat{\tau}^z_k 
+ \sideset{}{'}\sum_{\{k,l\}} Q_{abkl} \hat{\tau}^z_k \hat{\tau}^z_l + \cdots \, ,
\end{equation}
where the sites $a$ and $b$ are excluded in the primed sums.

The early-time growth of $C(t)$ shows the quadratic behavior indicated 
by the leading-order term of
\begin{equation} \label{eq:C_lbit_early}
C(t) = 8 \langle [\hat{J}^{\mathrm{eff}}_{ab}]^2 \rangle t^2 + O(t^4) \, .
\end{equation}
Within the effective Hamiltonian, this $t^2$ growth form of $C(t)$ may characterize 
the MBL systems and is qualitatively independent of a state prepared 
for the measurement and a particular disorder realization
as long as $\hat{J}^{\mathrm{eff}}_{ab}$ is essential.

The late-time behavior of $C(t)$ shows oscillations with a period and 
decay factor that depend on the disorder realization and the state used 
for measurement. In the disorder average, it is dephased and 
saturates around $C=1$~\cite{Chen2017,Swingle2017}. 
However, it is worth noting a particular type of the measurement 
with an eigenstate in which all moments of $\hat{J}^{\mathrm{eff}}_{ab}$ 
are the same, where the OTO commutator becomes a simple oscillation 
as $C_\mathrm{eig}(t) = 1 - \cos( 4t\langle \hat{J}^{\mathrm{eff}}_{ab}\rangle )$. 
In this special case, the $t^2$ growth is a transient behavior of 
the simple oscillation with a long period 
$\pi/2\langle \hat{J}^{\mathrm{eff}}_{ab}\rangle$ given by the effective 
interaction.

While the early-time $t^2$ behavior is persistent at any given disorder realization
in this phenomenological model, quantitative fluctuations across different realizations
of disorders can also be important. Here we briefly discuss the shape of 
the probability distribution $P(C)$ measured at a given early time $t$
over random disorder realizations for the later comparison with the results
in more realistic models.
For simplicity, let us borrow the uniform distribution 
$[ -2^{-l/2} e^{-l/\zeta}, 2^{-l/2} e^{-l/\zeta} ]$ from Ref.~\cite{Chen2017} 
for a random interaction term with the farthest-apart spins of distance $l$ 
in Eq.~\eqref{eq:Heff} to produce the effective interaction 
with a decay length $\zeta$. 
For the evaluation of $C(t)$, we consider the two particular types 
of the states that include the maximally mixed state 
$\hat{\rho}\propto \mathbb{1}$ at infinite temperature 
and the pure state given by an eigenstate.
In both cases, the central limit theorem works straightforwardly,
providing a Gaussian shape of $P(C)$ with the average and width 
increasing as $t^2$ for the case of the maximally mixed state
and $P(C) \sim C^{-1/2}\exp{(-C/c_t)}$ with the cutoff $c_t$ increasing 
as $t^2$ for the case of an eigenstate.

The measurement at infinite temperature seems to be ideal in the sense that 
it gives the Gaussian distribution centered at the disorder average moving as $t^2$. 
However, for a general state, the location of the average may not indicate 
a typical value as exemplified in the heavy-tailed distribution in the evaluation 
with an eigenstate. Nevertheless, in the phenomenological model, one can obtain 
the qualitatively correct growth of $C(t)$ from the disorder average, 
no matter how severe the fluctuations are, since every disorder realization 
produces the same early-time $t^2$ behavior starting from $t=0$.

This ideal situation that allows $C(t) \propto t^2$ to be observed 
at an arbitrarily early time implies that the particular process of 
the scrambling is the only dynamics in the l-bit model. 
In more realistic MBL systems, it is reasonable to 
assume the presence of other system-specific dynamics 
that may coexist with, or perhaps obscure, the characteristic 
scrambling behavior expected. Thus, it is a nontrivial question 
to ask whether or not one can see in practice the $t^2$ behavior
beyond the effective Hamiltonian.
In the next sections, we will present how the OTO commutator growth
develops in the disordered XXZ and mixed-field Ising models.

\section{Random-Field Heisenberg XXZ Model} \label{sec:XXZ}

\begin{figure*}[t]
\includegraphics[width=0.99\textwidth]{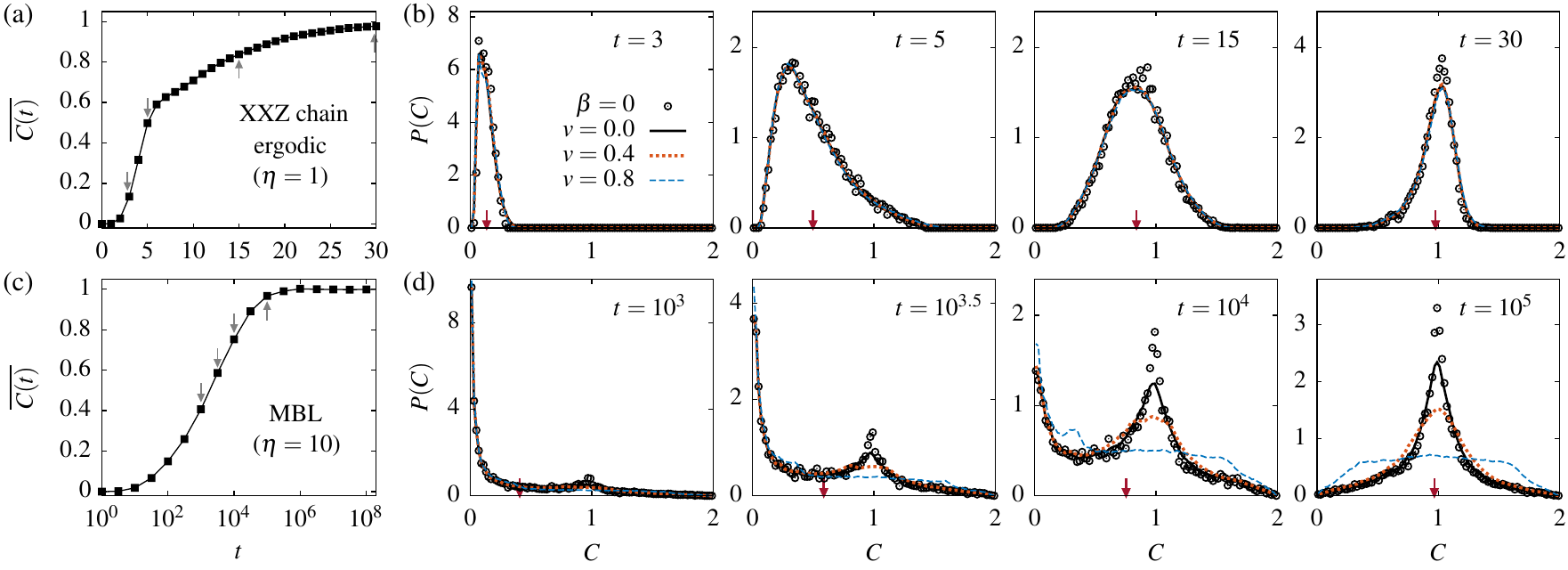}
\caption{Probability distribution of the OTO commutator measured 
along the growth in the random-field Heisenberg XXZ chain. 
The disorder average $\overline{C(t)}$ and the distribution $P(C)$ 
estimated at selected times over the disorder realizations 
are displayed for the disorder strengths of $\eta = 1$ 
and $\eta = 10$ in the ergodic (a, b) and MBL (c, d) phases, respectively. 
The OTO commutator $C$ is measured with the maximally 
mixed state at infinite temperature ($\beta=0$) and the random pure states 
at several values of $v$. 
The locations of the disorder average are indicated by the arrows 
in (b) and (d) which are not affected by the choice of the state. 
The distributions are estimated with $10000$ disorder realizations 
and averaged over $100$ random states at each $v$ in the system of length $L=12$.
} 
\label{fig1}
\end{figure*}

We first consider the spin-$1/2$ Heisenberg XXZ chain 
with a random Zeeman field given by the Hamiltonian, 
\begin{equation} \label{eq:Hxxz}
    \mathcal{H} = -\sum^{L-1}_{i=1} \left[ \hat{S}^x_i \hat{S}^x_{i+1} 
    + \hat{S}^y_i \hat{S}^y_{i+1} + J_z  \hat{S}^z_i \hat{S}^z_{i+1}\right] 
    +\sum_{i=1}^L h_i \hat{S}^z_{i} \, ,
\end{equation}
where the random field $h_i$ is uniformly sampled from the range of 
$[-\eta,\eta]$. The energy unit and $\hbar$ are set to be unity.
It is known that many-body localization occurs in this system 
for nonzero interactions $J_z \neq 0$ and strong enough disorder strength $\eta$
\cite{Pal2010,Znidaric2008,Bardarson2012,DeLuca2013,Luitz2015}. 
Here we fix the interaction at $J_z=0.2$ and consider the two particular  
values of the disorder strength, $\eta=1$ and $\eta=10$, where the system 
belongs to the ergodic and MBL phases, respectively. 
We define the OTO commutator $C(t)$ by choosing the Pauli spin operators
as $\hat{W}=\hat{\sigma}^x_4$ at site $i=4$ and $\hat{V}=\hat{\sigma}^x_1$ 
at one end of the chain. We employ the exact diagonalization 
for the numerical calculations of $C(t)$ in the systems of length $L=12$.

For the measurement of $C(t)$, we mainly consider the maximally mixed state 
$\hat{\rho} = \mathbb{1}/2^L$ at infinite temperature but also examine 
the random pure state $\hat{\rho} = |\Psi_v\rangle\langle\Psi_v|$ given by 
\begin{equation} \label{eq:bloch}
    |\Psi_v \rangle = \bigotimes_{i=1}^L 
    \left( \cos \frac{\theta_i}{2} |\uparrow\rangle
    + e^{i \phi_i} \sin \frac{\theta_i}{2} |\downarrow\rangle \right)
\end{equation} 
which is a product of local spin states sampled on the Bloch sphere.
We follow the scheme of Ref.~\cite{Nanduri2014} where
$\phi_i$ is a uniform random variable in $[0,2\pi)$, and 
the polar angle $\theta_i$ is restricted for $\cos\theta_i$ 
to be either $v$ or $-v$ at each site.
At $v=0$, $|\Psi\rangle$ is the sum of all $\sigma_z$-basis vectors 
with random phases, sharing the constant diagonal part of $\hat{\rho}$
with the maximally mixed state.
On the other hand, $v=1$ selects one of the $\sigma_z$-basis vectors 
that would be locally similar to an eigenstate in the strong-disorder limit.
Thus, varying $v$ may provide a systematic way to demonstrate the dependence of 
the estimate of $C(t)$ on the state preparations.  

We find that the probability distribution $P(C)$ of the OTO commutator 
shows contrasting behavior between the ergodic and MBL phases of 
the disordered XXZ chain. Figure~\ref{fig1} presents the time evolution 
of $P(C)$ along the main growth of the commutator measured 
over $10000$ random disorder realizations.
In the ergodic phase, a singly peaked distribution moves from $C=0$ to $1$ 
as the time goes, and its shape is independent of the parameter $v$ of 
the random pure states used for measurement, which also agrees 
well with the shape observed with the infinite-temperature state.

In contrast, in the MBL phase, the distribution $P(C)$ turns out to be 
bimodal at the intermediate times of the main growth in the measurement 
with the infinite-temperature state and random pure states with small $v$.
In the time evolution of $P(C)$, the population migrates from one peak 
at $C \approx 0$ to the other at $C \approx 1$.
The shape depends on the parameter $v$ since the late-time saturation 
behavior varies with $v$. The disorder average of $C(t)$ does not depend 
on the choice of the state, which, however, does not imply 
that the disorder average presents a typical behavior of $C(t)$.
The bimodal distributions observed at the intermediate times of 
the growth indicate that the disorder average is not physically 
meaningful in this regime. Indeed, the growth of the disorder average
is quite different from the characteristic behavior observed 
at an individual disorder realization which we present below. 

\begin{figure}
\includegraphics[width=0.48\textwidth]{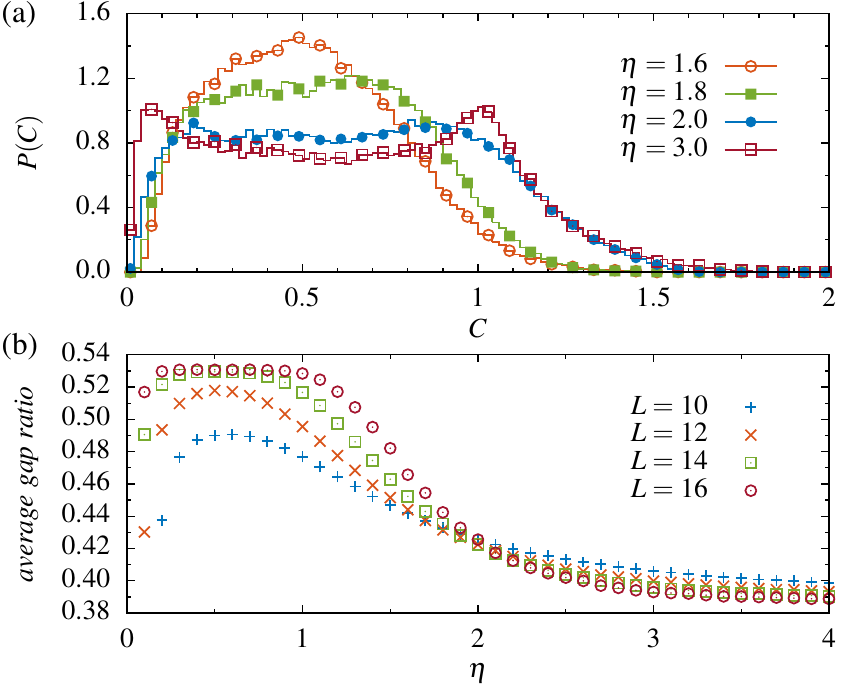}
\caption{OTO commutator distribution
around the localization transition in the XXZ chain. 
(a) Distribution $P(C)$ measured at $\overline{C(t)} \approx 0.6$
with the infinite-temperature state.
(b) Average gap ratio computed for eigenvalues $E$
within the distance $\Delta e = 0.1$ from the spectrum center
$e = 0.5$, where $e \equiv (E - E_\mathrm{min})/(E_\mathrm{max} - E_\mathrm{min})$,
and $E_{\mathrm{max}(\mathrm{min})}$ is the largest (smallest) eigenvalue.
}
\label{fig2}
\end{figure}

The emergence of the bimodality in $P(C)$ may work 
as an empirical indicator of the localization transition 
in the XXZ chain. Figure~\ref{fig2} displays the transition
in the shape of $P(C)$ with the disorder strength $\eta$ 
examined with the average given at $\overline C \approx 0.6$.
As $\eta$ increases, the top of the unimodal distribution 
in the ergodic phase becomes flat, and then double peaks start
to develop at a larger $\eta$. 
It turns out that the range of $\eta$ in which the transition
in the distribution shape occurs is consistent 
with the area of the localization transition indicated 
by the average gap ratio~\cite{Oganesyan2007,Luitz2015}
which we evaluate for our parameter $J_z = 0.2$ 
in Fig.~\ref{fig2}(b).

Since $P(C)$ in the MBL phase of the XXZ model is very different 
from the Gaussian distribution found in the effective l-bit 
model, natural questions are then what are the origins of 
the bimodal distribution and how it is related to the slow 
characteristic dynamics of $C(t)$ found in the effective model.
In order to address these questions, we look into the growth behavior 
of $C(t)$ at the level of individual disorder realizations. 
It turns out that the early-time stage of $C(t)$ is distinguished from 
the main growth stage, giving a waiting period at a very small $C$ 
after which the main growth starts to become visible. 
The time period of each growth stage shows large deviations between 
different disorder realizations, contributing to the slowly 
decreasing population at $C \approx 0$ and the broad distribution
over intermediate values of $C$ with another peak appearing 
due to the saturation around $C \approx 1$.

\begin{figure}
\includegraphics[width=0.48\textwidth]{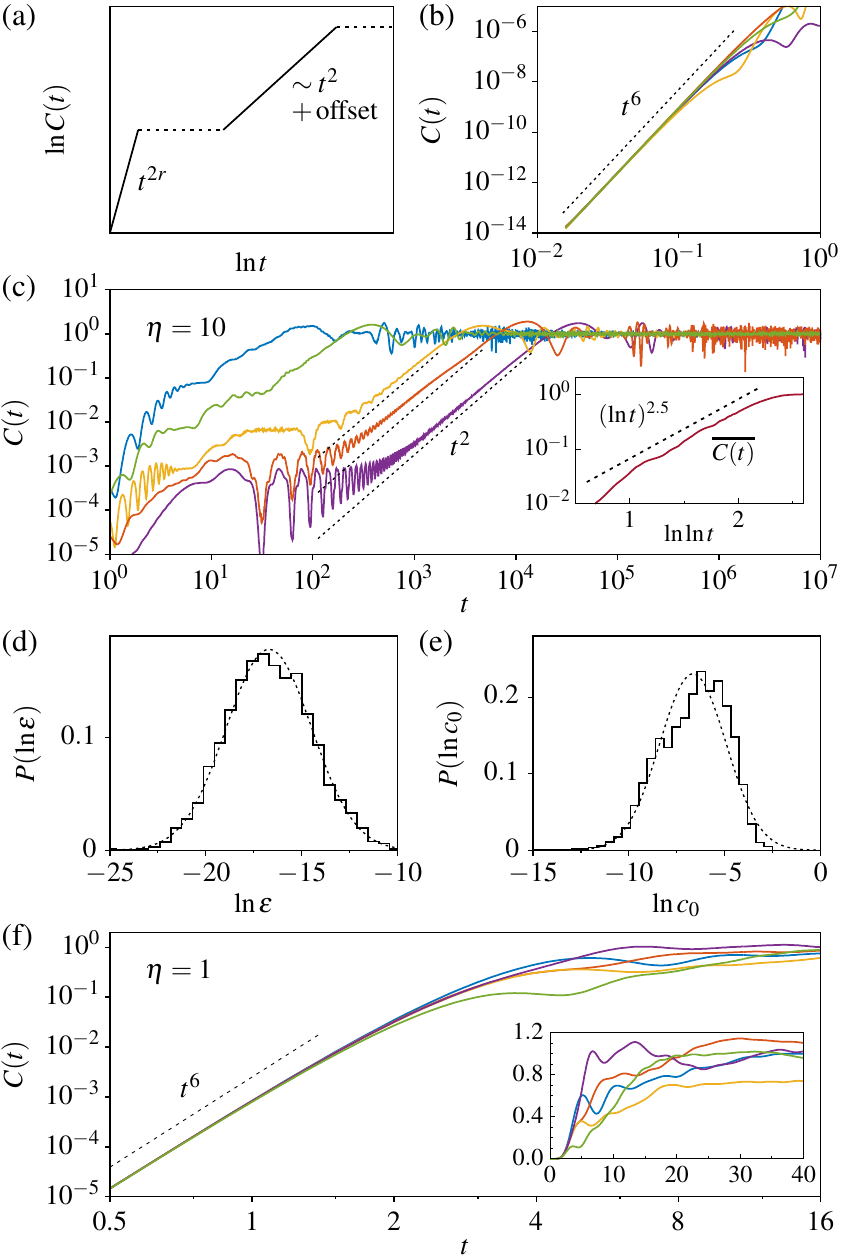}
\caption{Growth of the OTO commutator in the MBL phase of the XXZ chain. 
(a) Schematic of the power-law behaviors (solid lines) and oscillatory 
relaxations (dotted lines) observed at an individual disorder realization.
Examples of five selected disorder realizations are shown 
for (b) the early-time and (c) intermediate-time growth of $C(t)$
at $\eta=10$ measured with the infinite-temperature state
in the system of $L=12$. 
The inset of (c) shows the disorder average $\overline{C(t)}$. 
The distributions of (d) $\epsilon$ and (e) $c_0$ from the fits of 
$C(t) = c_0 + \epsilon t^2$ to the main growth parts 
are compared with the log-normal ones (dotted lines) 
having the same averages and variances.
(f) Examples of $C(t)$ at $\eta=1$ in the ergodic phase 
given for comparison.
}
\label{fig3}
\end{figure}

While the disorder average at a given time is largely influenced 
by the populations of $C \approx 0$ and $C \approx 1$, the main growth 
behavior observed at an individual disorder realization shows 
the characteristic feature expected from the effective model. 
Figure~\ref{fig3}(a) schematically describes the behavior observed 
at each stage.
At very early times, $C(t)$ shows the initial power-law growth behavior 
of $t^6$ for our choice of the two local operators with distance $r=3$.
The initial growth is relaxed shortly at the level of small $C$,
leading to the oscillatory relaxation plateau. The main growth emerges 
with another yet characteristic power-law type behavior, which becomes
saturated with long-period oscillations at late times.

The initial power-law growth is an intrinsic feature of the base spin model
and is not related to many-body localization. 
The early-time behavior can be easily understood from the series expansion 
of the commutator $[ \hat{\sigma}^x_{r+1}(t), \hat{\sigma}^x_1 ]$. 
Following the same procedures of Ref.~\cite{Dora2017}, 
the Baker-Campbell-Hausdorff expansion of $\hat{\sigma}^x_{r+1}(t)$ provides
\begin{equation}
    \hat{\sigma}^x_{r+1}(t) = \hat{\sigma}^x_{r+1} 
    + it [\mathcal{H}, \hat{\sigma}^x_{r+1}] 
    + \frac{(it)^2}{2!} [\mathcal{H}, [\mathcal{H}, \hat{\sigma}^x_{r+1}] ] 
    + \cdots \, ,
\end{equation}
where the first appearance of the term that does not commute with
$\hat{\sigma}^x_1$ is associated with $t^r$, and therefore, 
the squared commutator grows as $t^{2r}$ at very early times. 
While this initial power-law form has been derived and discussed 
previously as a general property of spin chains~\cite{Dora2017,Xu,Riddell2019},
it is still important to recall that the leading-order $t^{2r}$ term
is independent of the disorders and can appear without any contribution 
of the interaction that is essential to the MBL phase.

\begin{figure}
\includegraphics[width=0.48\textwidth]{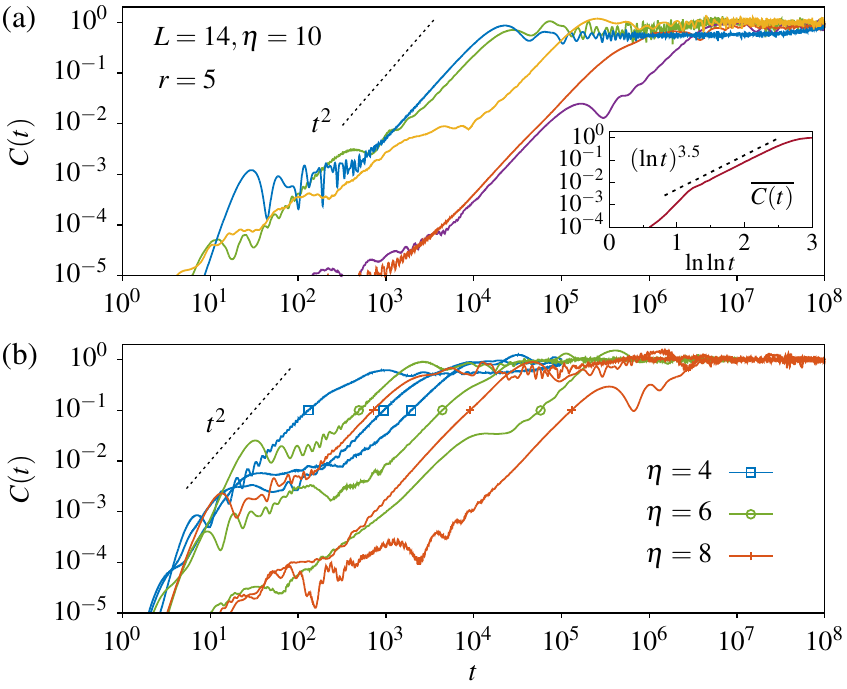}
\caption{Verification of the main power-law growth in a larger
system of the disordered XXZ chain.
The OTO commutator with the operator distance $r=5$ is considered 
in the systems of length $L=14$. (a) Five selected examples of
the $t^2$ growth behavior at $\eta=10$. The inset of (a)
indicates the disorder average $\overline{C(t)}$. 
(b) Examples of the $t^2$ behavior at weaker disorder strengths 
in the MBL phase.
}
\label{fig4}
\end{figure}

The second growth behavior of a power-law type that leads to a main 
increase in $C$ becomes visible after the plateau of oscillatory 
relaxations that suppress the initial $t^{2r}$ growth. 
Figure~\ref{fig3}(c) shows typical examples of the intermediate-time 
quadratic growth behavior at individual disorder realizations 
that are drastically different from the behavior of the disorder
average that looks like $\overline{C(t)} \sim (\ln t)^{2.5}$. 
The behavior of $\overline{C(t)}$ is not universal and depends on $r$ 
and the model systems as shown in Figs. \ref{fig4} and \ref{fig7}.
Unlike the effective model, the characteristic behavior can hardly
survive in the disorder average done at a given time that includes 
different stages of growth because of the large fluctuations 
of their time periods, and thus the quadratic growth is only 
identifiable in the level of individual disorder realizations. 

\begin{figure}
\includegraphics[width=0.48\textwidth]{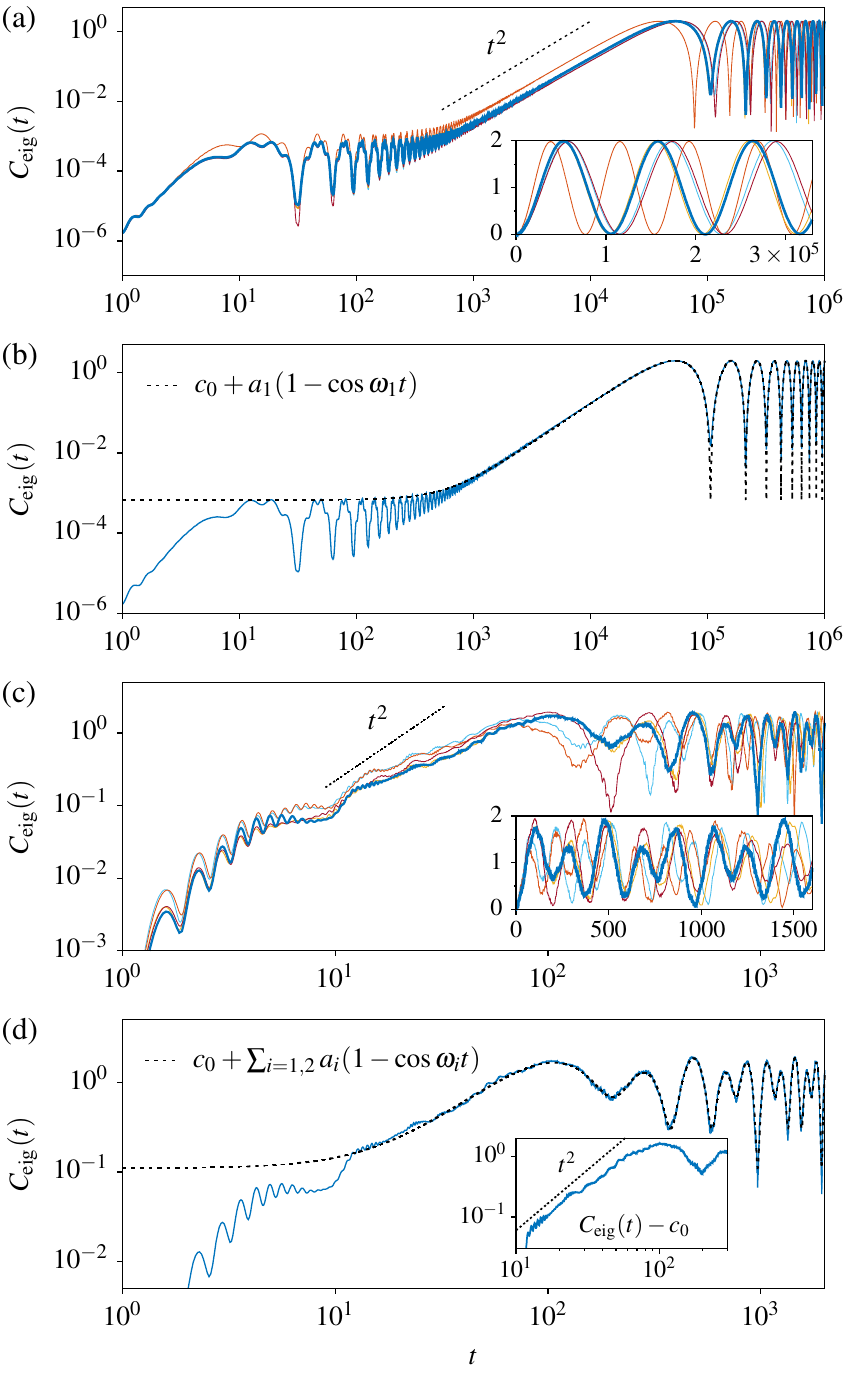}
\caption{OTO commutator measured with eigenstates at two given 
disorder realizations. The disorder realizations chosen for 
(a, b) and (c, d) are associated with the rightmost and leftmost lines 
in Fig.~\ref{fig3}(c), respectively, which contrast the cases of 
an excellent fit to the $t^2$ form and a poor power-law fit.
The lines displayed in (a) and (c) are 
different measurement with five eigenstates randomly chosen 
from the middle of the spectrum, 
among which one line is presented again in (b) and (d) 
for a fit to the empirical formula of the $(1-\cos \omega_i t)$ oscillations 
with an offset $c_0$. 
}
\label{fig5}
\end{figure}

While the intermediate-time growth often shows an excellent fit 
to the $t^2$ form found in the l-bit model, we find that the main growth 
is better characterized as $C(t) \approx c_0 + \epsilon t^2$ with 
an offset $c_0$ representing the contributions of short-time dynamics 
indicated by the relaxation plateau. 
A large value of the plateau makes a poor fit to the strict form of $t^2$ 
as can be seen in the examples given in Fig.~\ref{fig3}(c).
Several dynamic processes of different time scales may coexist in the OTO 
commutator. In the sense of a fixed-point Hamiltonian, the $t^2$ component 
may represent the long-time scrambling dynamics while the coarse-grained 
effects of all faster dynamics appear as a constant offset 
in the scale of intermediate and later times.

Figures~\ref{fig3}(d) and \ref{fig3}(e) display the distributions of $c_0$ 
and $\epsilon$ in the logarithmic scale. In particular, the one
for $\epsilon$ contrasts with the Gaussian distribution of the corresponding
quantity $\langle [\hat{J}^\mathrm{eff}]^2 \rangle$ in the l-bit model.
Given that $\langle [\hat{J}^\mathrm{eff}]^2 \rangle$ was an addition 
of random variables at the maximally mixed state, the log-normal 
shape could suggest that $\epsilon$ would come from a multiplicative process 
with random disorders, characterizing the large deviations between individual 
disorder realizations.

Figure~\ref{fig4} verifies the power-law behavior of $C(t)$ 
for a longer operator distance $r=5$ in the system of $L=14$ 
while we have mainly considered $r=3$ and $L=12$.
The power-law behavior is demonstrated at several values of 
the disorder strength $\eta$ of the MBL side. 
The time scale of the main growth stage tends to increase 
with $r$ and $\eta$, which influences the visibility of 
the power law in a finite system. At a weak $\eta$ close to 
the transition, one would need a very large $r$ for 
the main growth time scale to become separated from the early-time 
processes. The $t^2$ growth is thus more pronounced in practice
at stronger $\eta$ in a system with small $r$,
and thus we mainly consider the deep MBL regime 
at $\eta=10$ in the XXZ chain.

A pure-state measurement with an eigenstate at a given disorder realization 
may demonstrate more clearly the essential role of the slowest dynamics 
in the MBL phase to the power-law growth at intermediate times and 
the long-time behavior appearing at late times. 
For a given eigenstate $|\alpha\rangle$, one can express the OTO commutator 
$C_\mathrm{eig}(t)$ explicitly as 
\begin{equation}
    C_\mathrm{eig}(t) = 1 - \mathrm{Re} \Big[
    \sum_{\beta,\gamma,\delta} 
    e^{it(E_\alpha - E_\beta + E_\gamma - E_\delta)}
    s_{\alpha\beta\gamma\delta} \Big] \, ,
\end{equation}
where $s_{\alpha\beta\gamma\delta} = 
\langle \alpha | \hat{\sigma}^x_3 | \beta \rangle
\langle \beta | \hat{\sigma}^x_0 | \gamma \rangle
\langle \gamma | \hat{\sigma}^x_3 | \delta \rangle
\langle \delta | \hat{\sigma}^x_0 | \alpha \rangle$, and
$E_\alpha$ is the energy of the eigenstate $|\alpha\rangle$.
Inspired by the effective Hamiltonian, the long-period oscillations 
in the MBL phase may be determined by a few smallest magnitudes 
of  $(E_\alpha - E_\beta + E_\gamma - E_\delta)$  with a dominant
$s_{\alpha\beta\gamma\delta}$. Then, one can 
write down the late-time expression of $C_\mathrm{eig}(t)$ as
\begin{equation} \label{eq:c_eig}
    C_\mathrm{eig}(t) \approx c_0 
    + \sum_{i=1}^n c_i (1 - \cos \omega_i t) \,,
\end{equation}
if $\omega_i$ is well separated from the larger
frequencies that are coarse-grained in an offset $c_0$ for the low-resolution
description at intermediate and late times. In $C_\mathrm{eig}(t)$, 
the $t^2$ form appears in a transient behavior of the cosine term 
as found in the effective l-bit Hamiltonian with an eigenstate.

Figure~\ref{fig5} demonstrates the existence of dominant long-period modes
by providing the two particular examples of the best and worst fits 
to the strict $t^2$ form chosen among the cases shown in Fig.~\ref{fig3}(c).
In the one with the best fit, it turns out that $C_\mathrm{eig}(t)$ 
at intermediate and late times can be described with just one frequency 
in Eq.~\eqref{eq:c_eig}. In the other one, the contributions of 
two frequencies are dominant with a relatively high offset, 
which has caused the poor fit to the strict power-law form at 
intermediate times if the offset is not considered.
In both cases, we have not observed qualitative variations with 
different choices of an eigenstate while the value of the frequency  
depends on the eigenstate. The number of dominant modes and the separation
between the frequencies depend mainly on the disorder realization.

The intermediate-time growth behavior, $C(t) \sim c_0 + \epsilon t^2$, 
observed at an individual disorder realization remains the same 
in all our choices of the state for measurement. 
We have examined the random pure states at various values of $v$. 
The one at $v=0$ is essentially the same as $C(t)$ obtained 
with the maximally mixed state at the infinite temperature 
while the one at $v=1$ is very similar to the measurement with
an eigenstate. These will be shown explicitly in the next section, 
where the observation of the characteristic behavior in the MBL phase
is verified in the mixed-field Ising model.

\section{Mixed-Field Ising Model} \label{sec:Ising}

Another MBL system that we consider is the random-transverse-field 
quantum Ising chain in a uniform longitudinal field~\cite{Chen2017}. 
The Hamiltonian is given as
\begin{equation}
    \mathcal{H} = -\sum_{i=1}^{L-1} \hat{\sigma}^z_i \hat{\sigma}^z_{i+1} 
    - \sum_{i=1}^L h_i \hat{\sigma}^x_i
    - h_z \sum_{i=1}^L \hat{\sigma}^z_i \,,
\end{equation}
where the random transverse field $h_i$ is sampled from the uniform 
distribution in $[-W,W]$. The strengths of the random field and 
the uniform longitudinal field are fixed at $W=10$ and $h_z=0.1$, 
respectively. The chain length $L=12$ is considered. The OTO commutator
is defined with the operators $\hat{W} = \hat{\sigma}^z_3$ and 
$\hat{V} = \hat{\sigma}^z_0$ by following Ref.~\cite{Chen2017}.
The maximally mixed state and random pure states are considered for
measurement, where the pure state is defined analogously as 
\begin{equation} \label{eq:bloch2}
    |\Psi_v \rangle = \bigotimes_{i=1}^L 
    \left( \cos \frac{\theta_i}{2} |x;+\rangle
    + e^{i \phi_i} \sin \frac{\theta_i}{2} |x;-\rangle \right)
\end{equation}
with the basis $|x;\pm\rangle$ of the Pauli operator $\hat{\sigma}^x$ 
along the axis of the random transverse fields.

\begin{figure}
\includegraphics[width=0.48\textwidth]{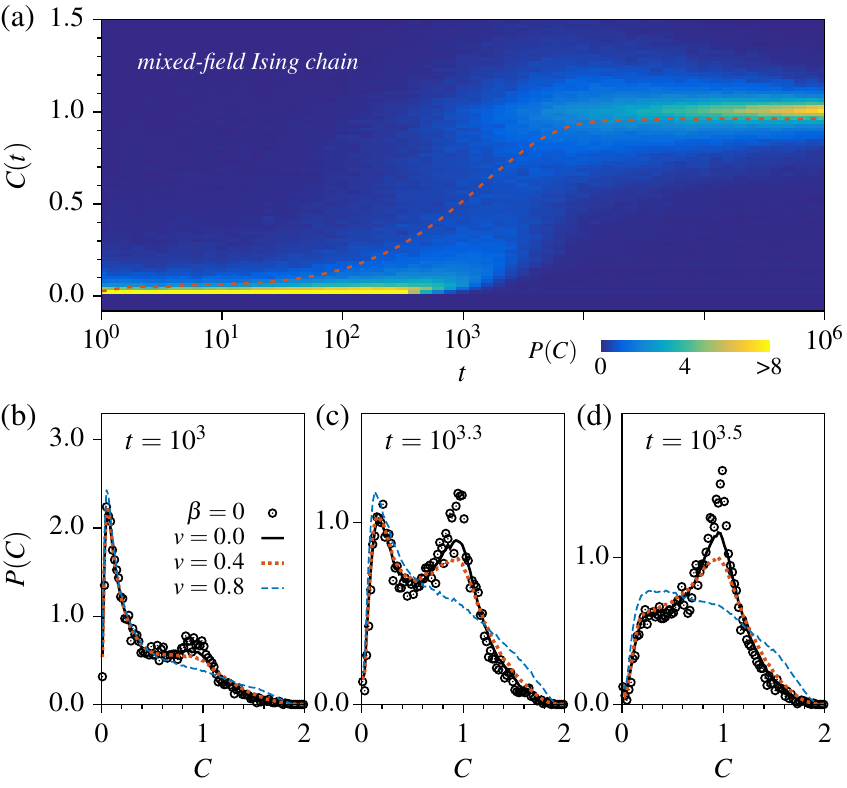}
\caption{Probability distribution of the OTO commutator in the mixed-field 
Ising chain. (a) The distribution $P(C)$ is visualized with 
the color code for the probability density of the OTO commutator $C(t)$ 
measured at a given time $t$ across $10000$ disorder 
realizations with the infinite-temperature state ($\beta=0$). 
The dotted line indicates the disorder average $\overline{C(t)}$. 
The random pure-state estimates for $v=0, 0.4, 0.8$ are compared with 
the infinite-temperature estimate in (b-d) at selected times 
along the main growth of $C(t)$.}
\label{fig6}
\end{figure}

As we have seen in the MBL phase of the XXZ chain, the disorder average 
$\overline{C(t)}$ in the mixed-field Ising chain also fails to represent 
the growth behavior of the OTO commutator $C(t)$ as indicated 
by Fig.~\ref{fig6}(a). The probability distributions of $C(t)$ 
measured across the random disorder realizations are very similar to 
those that we have observed in the XXZ chain.
The main growth of $C(t)$ appears with a time delay that fluctuates 
significantly across the disorder realizations, which leads to 
large deviations in the distribution $P(C)$. In particular,
at intermediate times, $P(C)$ shows a clear double-peak structure 
when measured with the infinite-temperature state and 
the random pure state at $v=0$, which agrees well with our previous 
observation in the XXZ chain.

\begin{figure}
\includegraphics[width=0.48\textwidth]{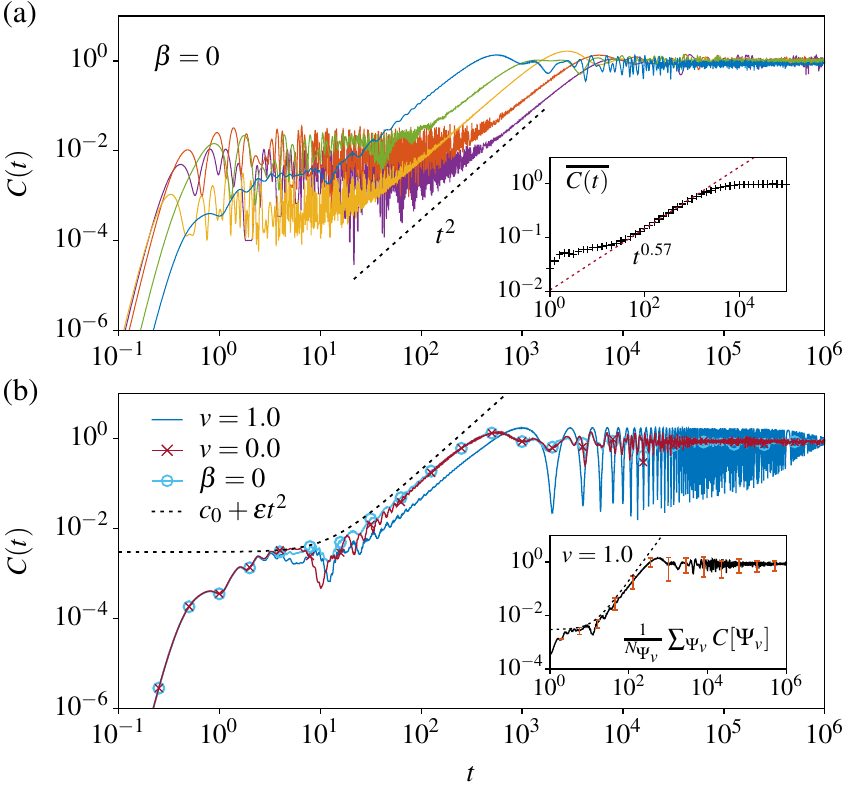}
\caption{OTO commutator in the mixed-field Ising chain at an individual 
disorder realization. (a) The growth of $C(t)$ of the infinite-temperature 
estimate ($\beta=0$) is plotted for five selected disorder realizations. 
The inset of (a) shows the disorder average $\overline{C(t)}$ calculated 
over $10000$ realizations for comparison. 
(b) The random-pure-state estimate is presented for $v=0.0$ and $1.0$
at a fixed disorder realization chosen from (a). 
The inset of (b) presents the averages (solid lines) and standard 
deviations (error bars) measure over $100$ generations 
of the random states drawn at $v=1$.
}
\label{fig7}
\end{figure}

The characterization of the growth behavior of $C(t)$ at an individual 
disorder realization shows excellent agreement with the multiple stages 
observed in the MBL phase of the disordered XXZ chain.
Figure~\ref{fig7} presents typical examples that show the initial 
power-law growth, the plateau of oscillatory relaxations,
the main growth of a power-law form at intermediate times, 
and then the saturation around $C \approx 1$ in the late times.
In the expansion of the commutator in the small time limit, 
one can easily show that the initial growth of $C(t)$ follows 
an intrinsic form of $C(t) \sim t^{4r+2}$, where $r=2$ for our choice of 
$\hat{W}$ and $\hat{V}$. A slight difference from the XXZ model 
is that the random transverse fields contribute to the $t^{4r+2}$ 
term, but it is still independent of the longitudinal 
field $h_z$ that triggers many-body localization in this system.

The intermediate-time main growth behavior also verifies the characterization 
of the power-law form $C(t) \sim c_0 + \epsilon t^2$ with an offset $c_0$
that is consistently observed in the different state preparations of 
the infinite-temperature state and the random pure states.
Averaging over disorder realizations distorts the main growth behavior 
as seen in the inset of Fig.~\ref{fig7}(a) because of the large deviations 
in the time period and level of the relaxation plateau setting the offset.
On the other hand, at a fixed disorder, our tests with different random 
states indicate that the main growth behavior of the power-law form 
is very robust in the intermediate-time regime. 

At a fixed disorder, the deviations between different random generations 
of the pure state appear mainly in the saturation stage at late times. 
For instance, in the case of $v=1$, the $(1-\cos\omega t)$ oscillations 
survive for a long period of time because of the significant participation 
of a single eigenstate. While these long-period oscillations are 
dephased effectively by averaging over many random states at the same $v$, 
the well-defined frequency of the oscillations at late times can 
be used for measuring an effective interaction. 
For a strong-disorder field, a single random pure state at $v=1$ 
generated in the axis of the disorder field may provide an approximate 
estimate of $\langle \hat{J}^{\mathrm{eff}} \rangle \propto \omega$ 
between the two local operators examined for the OTO commutator
growth behavior.

\section{Summary and Conclusions} \label{sec:summary} 

In conclusion, our results suggest that in the MBL systems, there exists 
a typical power-law-like behavior in the main growth of the OTO 
commutator at intermediate times that can be observed in the level of 
an individual disorder realization. The onset time of the main growth
shows large fluctuations across the disorder realizations, requiring 
the study of the OTO commutator growth to be constrained to each 
realization of disorders. For a fixed disorder configuration, 
the main growth characteristics at intermediate times are unaffected 
by different choices of the maximally mixed or random pure states 
prepared for the measurement.

We have examined two MBL systems of the Heisenberg XXZ chain with 
a random Zeeman field and the random-transverse-field Ising chain with
a uniform longitudinal field. At an individual disorder realization, 
both show qualitatively the same behavior of the OTO commutator $C(t)$. 
The initial growth exhibits an intrinsic power law that is unrelated 
to many-body localization. The initial growth is shortly suppressed 
by the relaxation plateau, making an offset added up to the $t^2$ behavior 
that appears as the main growth of $C(t)$. We have argued that 
the dominantly slow components of the OTO commutator found at an eigenstate 
is the essence of the $t^2$ growth behavior as indicated in the effective 
Hamiltonian, while the fast components are coarse-grained 
in the offset $c_0$ of $C(t) \sim c_0 + \epsilon t^2$ observed 
at intermediate times.

The disorder average often works as a convenient tool to
study a disordered system by reducing statistical noises. 
However, our calculations indicate that it may not generalize for $C(t)$ 
in the MBL systems. It turns out that the time scale and level of 
the relaxation plateau fluctuate severely across disorder realizations, 
preventing the disorder average from capturing the characteristic growth 
at intermediate times. Although the estimate of $C(t)$ may not be
self-averaging, the typicality of the power-law behavior that 
we have identified at individual disorder realizations raises 
a possibility that one may still be able to characterize MBL systems 
with the OTO commutator just by testing a few samples 
of quenched disorders.

\begin{acknowledgments}
This work was supported from the Basic Science Research Program 
through the National Research Foundation of Korea funded by 
the Ministry of Education (NRF-2017R1D1A1B03034669).
\end{acknowledgments}

\end{document}